\documentclass[a4paper,UKenglish,cleveref,autoref,thm-restate]{lipics-v2021} 
\pdfoutput=1

\usepackage{cite}
\usepackage{amsfonts}
\usepackage{algorithmic}
\usepackage{makecell}
\usepackage{url}
\usepackage{xurl}
\usepackage{tikz}
\usetikzlibrary{arrows.meta, positioning}
\usepackage[most]{tcolorbox}
\usepackage{booktabs}

\title{Forecasting the Maintained Score from the OpenSSF Scorecard: A Study of GitHub Repositories Linked to PyPI Packages}

\titlerunning{Forecasting the Maintained Score from the OpenSSF Scorecard}

\author{Alexandros Tsakpinis}
{fortiss GmbH, Munich, Germany}
{tsakpinis@fortiss.org}
{https://orcid.org/0000-0001-6561-2866}
{}

\author{Efe Berk Ergüleç}
{Technical University of Munich, Munich, Germany}
{efeberk.erguelec@tum.de}
{https://orcid.org/0009-0001-3629-2020}
{}

\author{Emil Schwenger}
{fortiss GmbH, Munich, Germany}
{schwenger@fortiss.org}
{https://orcid.org/0009-0009-6557-5095}
{}

\author{Alexander Pretschner}
{Technical University of Munich, Munich, Germany}
{alexander.pretschner@tum.de}
{https://orcid.org/0000-0002-5573-1201}
{}

\authorrunning{A. Tsakpinis, E. Ergüleç, E. Schwenger, and A. Pretschner}

\Copyright{Alexandros Tsakpinis, Efe Berk Ergüleç, Emil Schwenger, and Alexander Pretschner}

\ccsdesc[500]{Software and its engineering~Software libraries and repositories}

\keywords{Maintenance Activities, Open-Source Software, Time Series Forecasting}

\relatedversion{}

\acknowledgements{}

\hideLIPIcs 

\begin{document}
\nolinenumbers
\maketitle

\begin{abstract}
    \textbf{Background:}
    The OpenSSF Scorecard is widely used to assess the security posture of open-source software repositories, with the \textit{Maintained} metric serving as a key indicator of recent maintenance activities, helping users identify actively maintained projects and potentially abandoned dependencies. However, the metric is inherently retrospective, providing only a short-term snapshot based on the past 90 days of repository activity and offering no insight into the future. This limitation complicates risk assessment for developers and organizations that rely on open-source dependencies. 
    \textbf{Aims:}
    In this paper, we investigate the feasibility of forecasting future maintenance activities as captured by the OpenSSF \textit{Maintained} score. 
    \textbf{Method:}
    Focusing on 3,220 GitHub repositories linked to one of the top 1\% most central PyPI libraries, as ranked by PageRank, we reconstruct historical \textit{Maintained} scores over a three-year period and frame the problem as a multivariate time series forecasting task. We study four target representations: the raw \textit{Maintained} score (0--10), a bucketed score capturing low (0--2), moderate (3--7), and high (8--10) maintenance levels, the numerical trend slope between consecutive scores, and categorical trend types (downward, stable, upward). We compare a machine learning model (Random Forest) and a deep learning model (LSTM) using training windows of 3--12 months and forecasting horizons of 1--6 months. 
    \textbf{Results:}
    Our results show that future maintenance activity can be forecasted with meaningful accuracy, particularly when using aggregated representations such as bucketed scores and trend types leading to accuracies above 0.95 and 0.79. Notably, simpler machine learning models perform at least on par with deep learning approaches, suggesting that effective forecasting does not require complex architectures. 
    \textbf{Conclusions:}
    These findings demonstrate that predictive modeling can complement existing Scorecard metrics, enabling more proactive assessment of maintenance activities in open-source software ecosystems.
\end{abstract}
\section{Introduction}

Over the years, Open Source Software (OSS) has become a fundamental element of the product lifecycle, driven by commercial, engineering, and quality reasons~\cite{ebert2008open}. Currently, OSS components constitute 80--90\% of the code in commercial products, highlighting their widespread adoption in the software industry~\cite{pittenger2016open, oss2022}. A key subset of OSS is software libraries, also known as packages, which are essential to modern software development. These libraries enable developers to leverage pre-existing, well-tested functionality, eliminating the need to build everything from scratch~\cite{bauer2012structured}. After a library is integrated into a project it becomes a dependency~\cite{cox2019surviving}. However, OSS dependencies are not without risks. They can be affected by vulnerabilities (e.g., Log4j~\cite{log4j2021}) or targeted by supply chain attacks (e.g., LiteLLM~\cite{litellm}, XZ Utils~\cite{xz2024}) posing significant challenges to the software industry~\cite{decan2018impact}. When such issues arise, the community supporting the library typically acts quickly to release a fixed version~\cite{rahkema2022swiftdependencychecker}. However, it can happen that support for a library is discontinued or temporarily suspended, leaving projects vulnerable due to the absence of maintenance activities~\cite{bauer2012structured, raemaekers2011exploring, tsakpinis2023analyzing}. To address this, tools like the OpenSSF Scorecard have been developed to assess the maintenance status of OSS libraries. Widely adopted across software repositories~\cite{scorecard-action-dependents} and used by several industry-relevant projects~\cite{prominent-scorecard-users}, the OpenSSF Scorecard performs 18 security-related checks on a project’s repository. One of these checks, referred to as the \textit{Maintained} check, tracks the maintenance activities within the past 90 days following a formula-based scoring scheme~\cite{openssf_scorecard}. This check enables users to assess whether an OSS project is actively maintained and to flag potentially abandoned dependencies that may pose increased security and reliability risks.

Despite the utility of the \textit{Maintained} check, it has a notable limitation: the lack of predictive capabilities. The metric relies exclusively on recent repository activities---such as commits and issue updates---within the past 90 days, providing only a snapshot of current maintenance efforts. While historical repository activity is available, the \textit{Maintained} check itself does not model temporal patterns and is limited to a fixed 90-day observation window, thereby lacking mechanisms to anticipate future maintenance behavior. Although this information is valuable for assessing the immediate state of a repository, it falls short in offering insights into the sustainability of its maintenance over time. This absence of foresight creates challenges for developers, organizations, and security analysts who need to make informed decisions about integrating or continuing to rely on specific libraries. Without a clear understanding of future maintenance trends, it becomes difficult to mitigate risks associated with abandonment or declining support. To address this gap, we aim to investigate to what extent predictive models can forecast the maintenance activities of GitHub repositories measured by the OpenSSF \textit{Maintained} score. The following research questions~(RQs) have been formulated to guide this investigation:

\textbf{RQ1:} To what extent can the raw OpenSSF \textit{Maintained} score of GitHub repositories be forecasted using machine learning and deep learning models?

\textbf{RQ2:} To what extent can the raw OpenSSF \textit{Maintained} score, when grouped into discrete buckets, be forecasted using machine learning and deep learning models?

\textbf{RQ3:} To what extent can the trend slope (rate of change between consecutive OpenSSF \textit{Maintained} scores) be forecasted using machine learning and deep learning models?

\textbf{RQ4:} To what extent can the trend type (downward, stable, upward) of the OpenSSF \textit{Maintained} score be forecasted using machine learning and deep learning models?


To address these research questions, first we select a set of GitHub repositories. We focus on GitHub in this study due to its sustained popularity as a platform for hosting code of OSS libraries~\cite{eghbal2020working}. Given the highly imbalanced \textit{Maintained} score across repositories within an ecosystem~\cite{zahan2023openssf}, and to avoid artificially balancing the dataset, we focus on repositories linked to the most central libraries within an ecosystem. This choice naturally leads to a more balanced distribution, as shown by our experiments in Section~\ref{sec:time_series_anaylsis}, while also ensuring that the results are most applicable to the parts of the ecosystem where they matter most. We selected PyPI as the target ecosystem, as it is widely used in modern software development and exhibits strong library integration patterns~\cite{decan2016topology, abdalkareem2020impact, Octoverse2025}. To determine the importance of libraries, we used the PageRank score, which has proven effective in identifying central libraries in ecosystems~\cite{tsakpinis2024analyzing, tsakpinis2025analyzing, mujahid2021toward}. For the selected repositories, we crawled commit histories, issue information, and project metadata, converting the unstructured data into a machine-readable format. Based on these data, we represented the target variable in four forms: the raw \textit{Maintained} score (0--10), a bucketed representation capturing low (0--2), moderate (3--7), and high (8--10) maintenance levels, the trend slope between consecutive scores, and the corresponding trend type (downward, stable, upward). We compared two forecasting models: a machine learning model (Random Forest, RF) and a deep learning model (Long-Short-Term-Memory, LSTM). Each method was evaluated for prediction accuracy using training windows of 3--12 months and forecasting horizons of 1--6 months.


Our contributions can be summarized as follows:
(1) We reconstruct historical OpenSSF \textit{Maintained} scores for GitHub repositories linked to PyPI libraries over a three year period, enabling a longitudinal time series analysis of a metric that is typically available only as a single point-in-time snapshot.
(2) We formulate the prediction of future maintenance activities as a multivariate time series forecasting problem and study four target representations: the raw \textit{Maintained} score, a bucketed score capturing maintenance levels, the numerical trend slope between consecutive scores, and categorical maintenance trend types.
(3) We empirically evaluate the predictive performance of machine learning (RF) and deep learning (LSTM) models across varying training window lengths and forecasting horizons.
(4) We show that future maintenance activity can be forecasted with meaningful accuracy---particularly when using aggregated representations such as bucketed scores and trend types---and that simpler machine learning models perform at least on par with deep learning approaches.
\section{Background and Related Work}

\subsection{OpenSSF Scorecard}
\label{sec:openssf_scorecard}

The \textbf{Open} \textbf{S}ource \textbf{S}ecurity \textbf{F}oundation (OpenSSF), part of the nonprofit Linux Foundation, aims to secure OSS for the greater public good. In November 2020, it introduced the Scorecard project to automate security assessments by assigning a score to OSS repositories, reducing manual analysis. Practitioners can run the Scorecard on GitHub and GitLab repositories to evaluate 18 metrics, each scored from 0 to 10. A weighted average of the individual metrics is also presented as an aggregated score. This study focuses on the \textit{Maintained} metric, which measures active maintenance through commit and issue activity by collaborators, members, or project owners~\cite{zahan2023openssf}. Archived projects receive the lowest score, while projects with at least one commit per week over the past 90 days score highest. Partial scores are awarded for issue activities involving collaborators, members, or owners. This metric is valid only for projects older than 90 days, as younger projects require manual review to ensure quality~\cite{openssf_scorecard}.

Related work has extensively applied the OpenSSF Scorecard in various security related studies. Younis et al. utilized the Scorecard to assess software supply chain security risks in industrial control system protocols, identifying strengths, weaknesses, and areas for improvement. Their findings highlight high scores for the \textit{Maintained} metric, reflecting active community engagement~\cite{younis2023analyzing}. Similarly, Mounesan et al. analyzed container-related threats by examining security attributes of open-source dependencies in their software supply chains. Their results revealed a highly uneven distribution of \textit{Maintained} scores: 69\% of dependencies scored 0, while only 16\% achieved the maximum score of 10. However, as 65\% of these packages originated from sources like deb, npm, and Java archives, the findings may have limited applicability to other ecosystems, such as PyPI~\cite{mounesan2023exploring}. Supporting these observations, Zahan et al. analyzed the PyPI ecosystem using OpenSSF Scorecard metrics and found similarly skewed distributions for the \textit{Maintained} metric: 75.9\% of packages scored 0, while only 24.1\% achieved scores between 1 and 10~\cite{zahan2023openssf}. In another study, Zahan et al. investigated whether selected OpenSSF Scorecard metrics could predict the number of externally reported vulnerabilities. Among others, the \textit{Maintained} metric emerged as a key predictor~\cite{zahan2023software}. Another study identified inactive packages as signals within the npm ecosystem, suggesting these packages are more vulnerable to supply chain attacks, where attackers could exploit these signals to launch such attacks. The authors propose leveraging the \textit{Maintained} metrics from the OpenSSF Scorecard to evaluate a library's maintenance activity automatically~\cite{zahan2022weak}. Beyond these studies, prior work has shown that a substantial share of ecosystem packages appear abandoned or unmaintained, amplifying security and supply-chain risks~\cite{vaidya2021securityissueslanguagebasedsoftware}. Although many referenced studies leverage OpenSSF Scorecard metrics in their methodologies, none have attempted to forecast Scorecard metrics to provide additional insights into the future. Moreover, all studies collected OpenSSF scorecard metrics at a single point in time, limiting the ability to assess score distributions over extended time periods. Such temporal analysis would offer a more realistic perspective on maintenance activities within an ecosystem. Addressing these two gaps is a primary focus of our study.

\subsection{Multivariate Time Series Forecasting}
Time series forecasting involves predicting future values of a sequence based on past data from the same signal (univariate forecasting) or from multiple related signals (multivariate forecasting)~\cite{romeu2013time}. Over time, forecasting methods have evolved from traditional statistical approaches to advanced machine learning and deep learning methods~\cite{spiliotis2023time}. Forecasting plays a crucial role in various domains, including economics, finance, weather simulation, and resource allocation~\cite{li2024deep}. 
In the context of software engineering~(SE), time-series forecasting has been applied directly to repository activity signals. Recent research has proposed models for predicting abandonment of OSS repositories~\cite{xu2025predictingabandonmentopensource}, highlighting the need for proactive indicators of future maintenance activities. Prior work has also forecasted commit activity in OSS projects using statistical and fuzzy time-series models~\cite{saini2016fuzzy}, as well as probabilistic models of developer contribution behavior~\cite{decan2020gap}. Other studies have used time-series forecasting to support software project state assessment based on repository event histories~\cite{romanov2023time}. Forecasting approaches have also been applied to additional SE metrics, including bug resolution times in OSS repositories~\cite{aversano2025time}, CI build outcomes (pass/fail) modeled as sequence-based classification tasks~\cite{saidani2022improving}, runtime software metrics for proactive monitoring~\cite{di2024time}, and the future number of reported vulnerabilities~\cite{kalouptsoglou2022time}. Collectively, these works demonstrate that forecasting software engineering metrics can support data-driven decision-making in various contexts. However, forecasting maintenance-related metrics as measured by the OpenSSF \textit{Maintained} score and its temporal dynamics remains largely unexplored, which we address in this study.

Predicting the OpenSSF \textit{Maintained} score based on the features used in its formula falls under multivariate forecasting~\cite{romeu2013time}. When performed beyond a single time step, this setting is categorized into iterative and direct multi-step-ahead approaches, as illustrated in Figure~\ref{fig:forecasting_types}.

\begin{figure}[ht]
    \centering
    \includegraphics[width=0.6\textwidth, trim=20mm 20mm 20mm 940mm, clip]{figures/Forecasting_Types.pdf}
    \caption{Overview of multi-step-ahead forecasting types}
    \label{fig:forecasting_types}
\end{figure}

Multi-step-ahead iterative forecasting predicts multiple future values by using previously predicted values as input for subsequent predictions, while multi-step-ahead direct forecasting predicts multiple future values without relying on predicted values as inputs for further predictions. Depending on the context, some studies suggest that iterative approaches are more efficient~\cite{marcellino2006comparison}, while others advocate for direct forecasting~\cite{yin2019experimental}. In this study, we use the latter approach and limit the forecasting horizon to six months, as the accuracy generally declines with longer horizons~\cite{mendis2024multivariate}. A decrease in performance may also reduce the practical applicability and user trust, further supporting the decision to limit the forecasting horizon.

Our selection of the concrete methods within two model families---machine learning (ML) and deep learning (DL)---is informed by prior research~\cite{saini2016fuzzy, decan2020gap, romanov2023time}, which demonstrates that temporal patterns in repository activity can be modeled effectively using such techniques. Note, that the selected methods are not intended to represent the best-performing state-of-the-art approaches in each family, but rather serve as well-established choices to assess whether forecasting the raw OpenSSF \textit{Maintained} score, the bucketed \textit{Maintained} score, the trend slope between consecutive scores, and the trend type is feasible with a given method.
For ML, we selected Random Forest (RF), a widely used ensemble method that trains multiple decision trees and generates a median prediction from them. This technique reduces overfitting and improves generalizability. RF has demonstrated strong performance in multivariate forecasting, even for long-term predictions, surpassing methods like support vector machines and multivariate adaptive regression splines~\cite{hamidi2018application}.
For DL, we chose Long Short-Term Memory~(LSTM) networks, which are effective at capturing long-term dependencies in sequential data and modeling temporal dynamics~\cite{hochreiter1997long}. LSTMs have been successfully applied to multi-step forecasting of high-dimensional time series data, outperforming alternatives such as support vector regression~\cite{bakir2018commerce}. However, experiments showed that LSTM models required several minutes to train, while RF finished within seconds~\cite{mussumeci2020large}.  
In summary, we must balance trade-offs between fast-training ML models, potentially less effective on complex data, and more powerful but computationally expensive DL models.
\section{Study Design}
\begin{figure}[ht]
    \centering
    \includegraphics[width=0.65\textwidth, trim=2mm 2mm 2mm 2mm, clip]{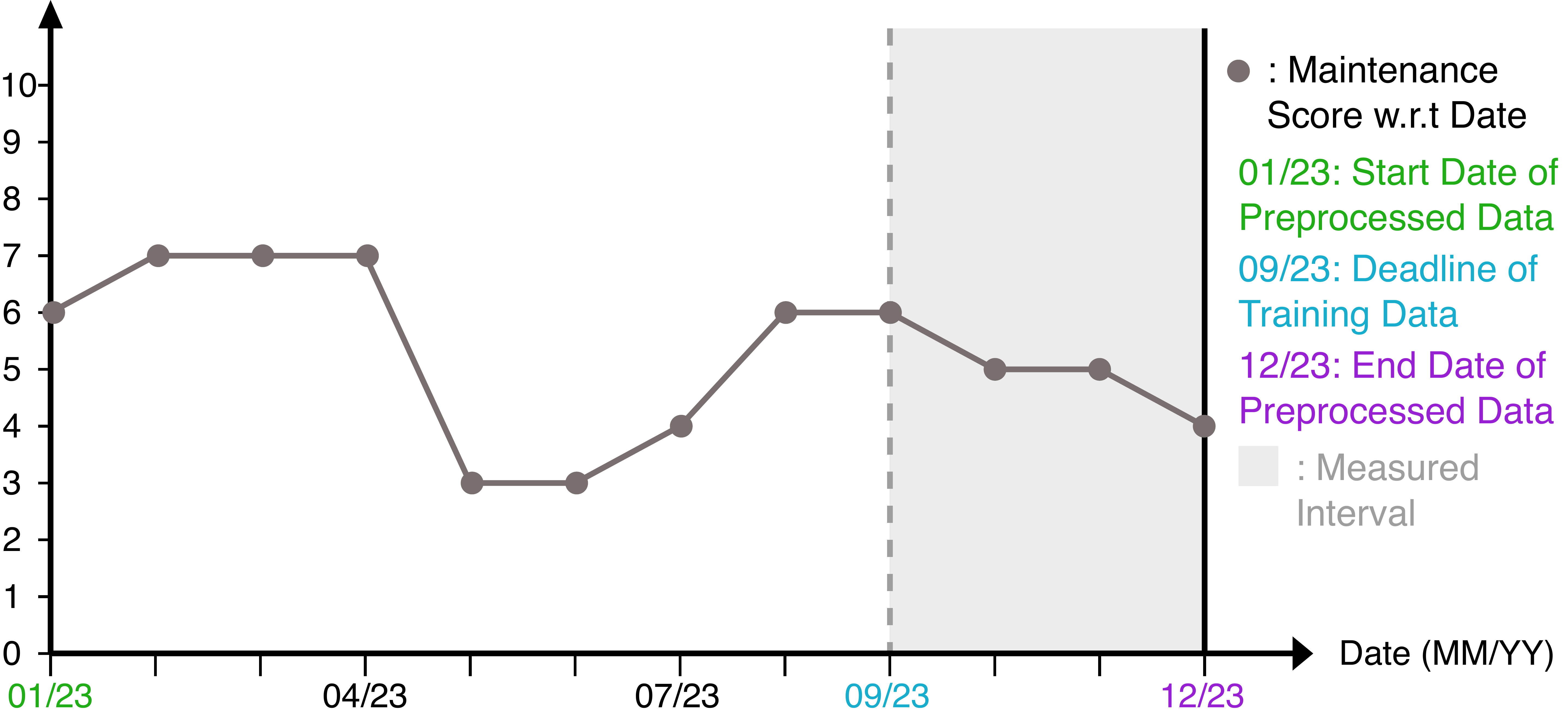}
    \caption{Forecasting the raw \textit{Maintained} score of the OpenSSF Scorecard (dates are exemplary)}
    \label{fig:scorecard_problem}
\end{figure}

We aim to forecast the raw OpenSSF \textit{Maintained} score (illustrated in Figure~\ref{fig:scorecard_problem}), the bucketed \textit{Maintained} score, the trend slope between consecutive scores, and the trend type using historical GitHub repository data across multiple forecasting horizons. To this end, we conduct a quantitative empirical study in which historical repository activity is reconstructed as multivariate time series and used to train and evaluate forecasting models under varying forecasting horizons. To ensure a reliable ground truth for the evaluation, we conduct the study retrospectively using data from the past.

\subsection{Data Collection and Preprocessing}
\subsubsection{GitHub Repository URL Collection}

As a first step, we defined a set of GitHub repositories for our experiments. We focused on repositories linked to libraries within a single ecosystem, as differences between ecosystems can vary significantly across multiple dimensions~\cite{zahan2023openssf} and thus limit generalizability. Therefore, we selected PyPI as the target ecosystem and used Python scripts to automate data collection from multiple sources. We first retrieved a comprehensive list of libraries via a provided endpoint~\cite{pypi-simple}, and then obtained detailed information---such as GitHub URLs and dependency data---through another endpoint~\cite{pypi-json}. For each GitHub URL, we verified accessibility and collected all direct and optional dependencies for the latest library version to construct a full dependency graph with unlimited transitive depth. This graph is used to filter repositories by relative importance, naturally leading to a more balanced dataset with respect to \textit{Maintained} scores. We focus on the top 1\% of PyPI libraries ranked by PageRank that include a valid GitHub link, resulting in 4,010 repositories. Further details on the characteristics of these repositories and the rationale for this selection are provided in Section~\ref{sec:time_series_anaylsis}.

\subsubsection{GitHub Repository Metadata Collection}
\label{sec:github_repo_metadata}

Based on these 4,010 GitHub repositories, we collect specific metadata to enable the recalculation of their \textit{Maintained} scores for a defined experimental period in the past. All 4,010 repositories are used for the time series analysis in Section~\ref{sec:time_series_anaylsis}, while a subset is later filtered for model training and evaluation, as outlined in Section~\ref{sec:representation_raw_score}. The \textit{Maintained} score formula requires various metadata, including project details (archived status, archived date, and creation date), issue-related data (issue creation date, issue creator role, associated comments, including comment creator roles and comment creation dates), and daily commit activity on the main branch. The metadata are collected from multiple sources: project and issue information from the GitHub GraphQL API, and commit history from the cloned repository. These metadata and their relationships are stored in a Neo4j graph database, and extracted as JSON using Python scripts and Cypher queries.

\subsubsection{Maintained Score Calculation}

Using the repository metadata collected in Section~\ref{sec:github_repo_metadata}, we reconstruct the daily \textit{Maintained} scores across the experiment period to obtain a consistent ground truth, i.e., the scores as they would be computed by the OpenSSF Scorecard based on historical repository states. The reconstruction follows the official specification and proceeds in three stages~\cite{openssf_scorecard}:

First, we derive daily activity signals from repository events. For each day $\tau$, let $C(\tau)$ denote the number of commits to the repository’s default branch and $I(\tau)$ the number of issue-related activities. Issue events include issue creations and comments authored by core roles only, namely \textsf{OWNER}, \textsf{MEMBER}, and \textsf{COLLABORATOR}. Duplicate entries per date and role are removed to avoid overcounting, and the resulting signals are expanded to a complete daily index with missing days assigned a value of $0$.
Second, activity is aggregated over an inclusive sliding 90-day lookback window. For each day $t$ in the evaluation period, we compute the aggregated commit activity (1) and issue activity (2), and a gating function (3):
\begin{align}
(1)\; S_c(t) &= \sum_{\tau=t-89}^{t} C(\tau) &
(2)\; S_i(t) &= \sum_{\tau=t-89}^{t} I(\tau) &
(3)\; g(t) &=
\begin{cases}
1, & t_{\text{start}} \le t \le t_{\text{end}},\\
0, & \text{otherwise}
\end{cases}
\notag
\end{align}
The gating function assigns a score of zero to archived repositories or those younger than $90$ days. Here, $t_{\text{start}}$ denotes the first day on which a repository becomes eligible for scoring (90 days after its creation), and $t_{\text{end}}$ denotes the archival date if present, otherwise the final day of the observation period. Third, the daily unrounded \textit{Maintained} score $\hat{M}(t)$ is computed using \texttt{LOOKBACK\_DAYS=90}, \texttt{ACTIVITY\_PER\_WEEK=1}, and \texttt{DAYS\_IN\_ONE\_WEEK=7}, together with the aggregated commit activity (1) and issue activity (2) within the 90-day window:
\[
\hat{M}(t)=10 \times 
\frac{S_c(t)+S_i(t)}
{\texttt{ACTIVITY\_PER\_WEEK}\cdot\left(\frac{\texttt{LOOKBACK\_DAYS}}{\texttt{DAYS\_IN\_ONE\_WEEK}}\right)}
\]
The final daily \textit{Maintained} score is obtained by applying the project-information gate (3), clipping the value to the maximum score of $10$, and rounding to the nearest integer:
\[
\text{Maintained}(t)=
\operatorname{round}\!\left(
g(t)\cdot\min\!\bigl(10,\hat{M}(t)\bigr)
\right)\in\{0,\dots,10\}
\]

\subsection{Time Series Analysis}
\label{sec:time_series_anaylsis}

The time series analysis provides insights that guided several design choices in our study. We conducted this analysis over the years 2021--2023, aligning with the period later used for model training and evaluation. This period was determined by the start of our study in early 2024. While extending the dataset to more recent months would have required substantial additional data collection effort without clear additional benefit, the selected window remains sufficiently recent to capture representative maintenance patterns.

\subsubsection{Empirical Evidence for Methodological Design Choices}

The following two investigations rely on the 4,010 repositories linked to one of the top 1\% of PyPI libraries ranked by PageRank and are based on the information reported in Table~\ref{tab:mean_activity}.

\textbf{Forecasting granularity:}
To identify an appropriate granularity for forecasting (daily, weekly, monthly, or yearly), we examined the average duration between consecutive commit and issue activities---specifically, issue creation and commenting events, which directly contribute to the calculation of the \textit{Maintained} score. Each year, over 3,000 repositories exhibited commit activity and over 2,400 showed issue activity, with partial overlap between the two groups. With mean intervals of 30--31 days between commits and issue-related activities, we selected \textbf{monthly} forecasting for subsequent model development and evaluation.

\textbf{Contributor stability:}
Next, we evaluated whether forecasting the \textit{Maintained} score is feasible based on the stability of repository activity---specifically, whether contributor behavior follows consistent and predictable patterns over time. We analyzed the monthly composition of contributors for each repository, considering code committers and issue participants (users who opened or commented on issues in core roles, i.e., \textsf{OWNER}, \textsf{MEMBER}, \textsf{COLLABORATOR}). Contributor overlap between consecutive months was quantified using the Jaccard similarity coefficient~\cite{travieso2024analytical, nunes2305dothash}, where higher values indicate recurring participation and stable engagement. Across the dataset, we observe over 1.75 million commits and approximately 900,000 issue interactions from around 61,000 unique participants. Commit activity exhibits moderate stability, with an average Jaccard similarity of 0.43--0.45, indicating that roughly 43--45\% of committers remain active across consecutive months. In contrast, issue activity shows high stability, with an average similarity of about 0.77, suggesting that a persistent group of maintainers consistently manages issues and discussions. This pattern---high stability in issue participation combined with moderate stability in commit activity---indicates that, although many code contributors participate only temporarily, a stable core of maintainers sustains continuous project and issue coordination. Overall, this temporal consistency supports reliable forecasting of maintenance activities.

\begin{table}[ht]
\centering
\caption{Yearly summary of repository activity and contributor stability}
\resizebox{\textwidth}{!}{%
\begin{tabular}{c|cc|cc|c||cc|cc}
\hline
\textbf{Year} &
\makecell{\textbf{Commit}\\\textbf{activity}} &
\makecell{\textbf{Active}\\\textbf{repositories}} &
\makecell{\textbf{Issue}\\\textbf{activity}} &
\makecell{\textbf{Active}\\\textbf{repositories}} &
\textbf{Mean} &
\makecell{\textbf{Commit}\\\textbf{activity}} &
\makecell{\textbf{Active}\\\textbf{repositories}} &
\makecell{\textbf{Issue}\\\textbf{activity}} &
\makecell{\textbf{Active}\\\textbf{repositories}} \\
\hline
2021 & 26 & 3063 & 34 & 2450 & 30 & 0.43 & 2776 & 0.77 & 2154 \\
2022 & 27 & 3118 & 34 & 2460 & 31 & 0.44 & 2812 & 0.77 & 2199 \\
2023 & 26 & 3112 & 34 & 2439 & 30 & 0.45 & 2779 & 0.77 & 2164 \\
\hline
\end{tabular}
}
\label{tab:mean_activity}
\end{table}

\subsubsection{Representation of the Target Variable as Raw Score}
\label{sec:representation_raw_score}

After establishing the temporal granularity and providing evidence for the feasibility of the study, we analyzed the distribution of \textit{Maintained} scores across the 4,010 repositories by examining their monthly values from 2021 to 2023. Prior work and our experiments report a highly skewed distribution of \textit{Maintained} scores in the PyPI ecosystem~\cite{zahan2023openssf, replication_package}. Instead of artificially re-balancing the data, we focus on the most central libraries ranked by PageRank. Restricting the analysis to the top 1\% captures the most impactful repositories while leading to a less extreme distribution. Within this subset, the remaining skew is still concentrated at the extremes. Thus, we exclude repositories that consistently exhibit scores of 0 or 10 from model training and evaluation, reducing the dataset from 4,010 to 3,220 repositories.

\begin{figure}[h]
    \centering
    \includegraphics[width=0.76\textwidth, trim=2mm 3mm 2mm 2mm, clip]{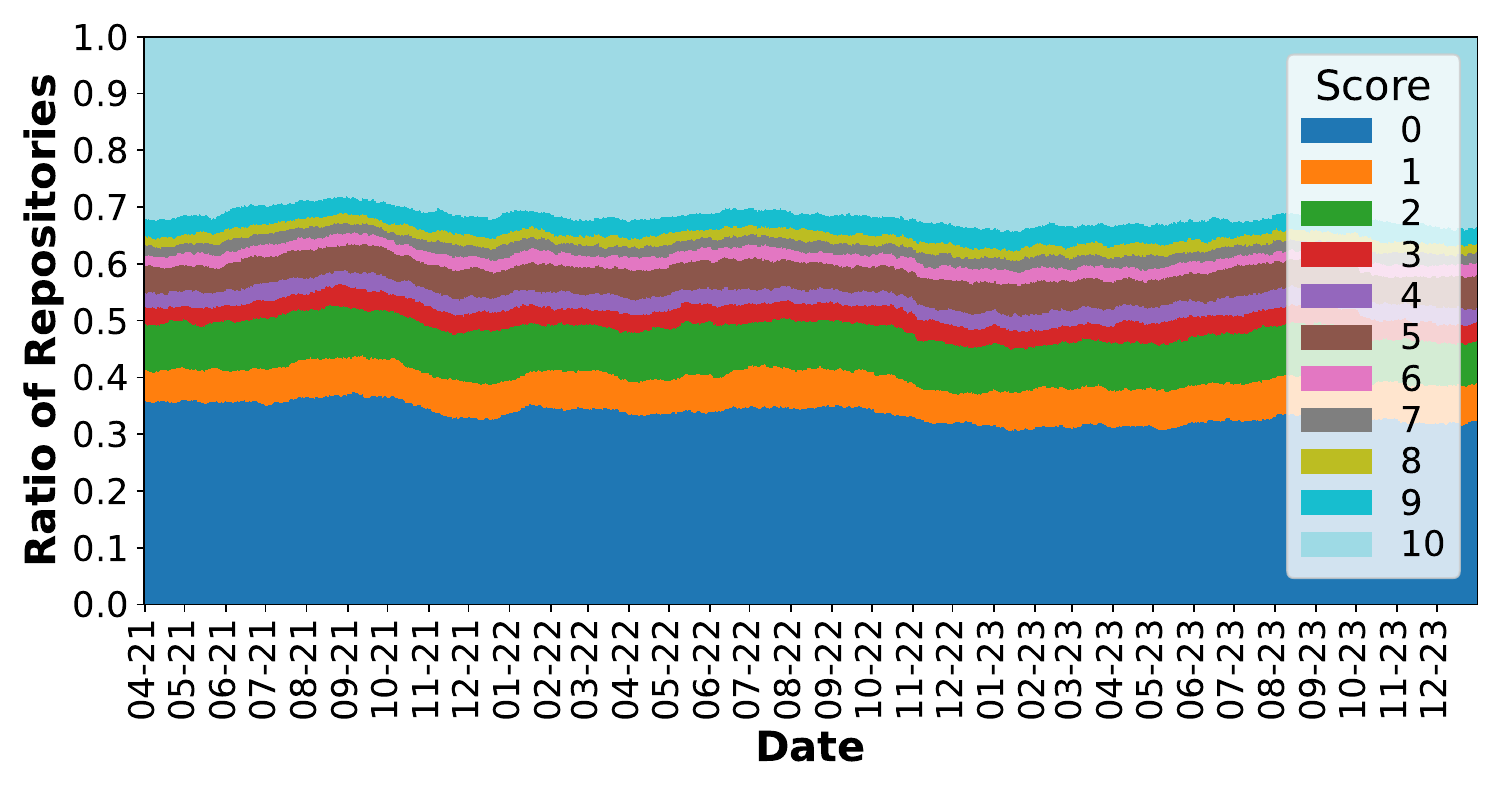}
    \caption{Temporal distribution of the raw \textit{Maintained} scores}
    \label{fig:rel_Distribution_Score_Time}
\end{figure}

Figure~\ref{fig:rel_Distribution_Score_Time} shows the temporal distribution of \textit{Maintained} scores between 2021 and 2023. The largest proportions occur at scores of 0 and 10, accounting for approximately 66\% of observations, while mid-range scores represent about 34\%. Compared to related work, the proportion of repositories with a score of 0 is reduced from 75.9\% to 34\%~\cite{zahan2022weak}. Overall, this results in a more balanced and temporally stable distribution of maintenance activity levels, providing a solid basis for subsequent forecasting tasks.

\subsubsection{Representation of the Target Variable as Bucketed Score}
While the raw \textit{Maintained} score provides a fine-grained view of repository activity, its distribution remains imbalanced, with most observations concentrated at the extremes (0 and 10). This limits the ability of forecasting models to effectively learn patterns across the full spectrum, particularly in the sparse mid-range. To address this, we aggregate scores into three buckets representing distinct maintenance levels: \textit{low}~(0--2), \textit{moderate}~(3--7), and \textit{high}~(8--10). This transformation smooths minor fluctuations in individual scores while preserving overall behavior leading to a more balanced target variable.

\begin{figure}[h]
    \centering
    \includegraphics[width=0.76\textwidth, trim=2mm 3mm 2mm 2mm, clip]{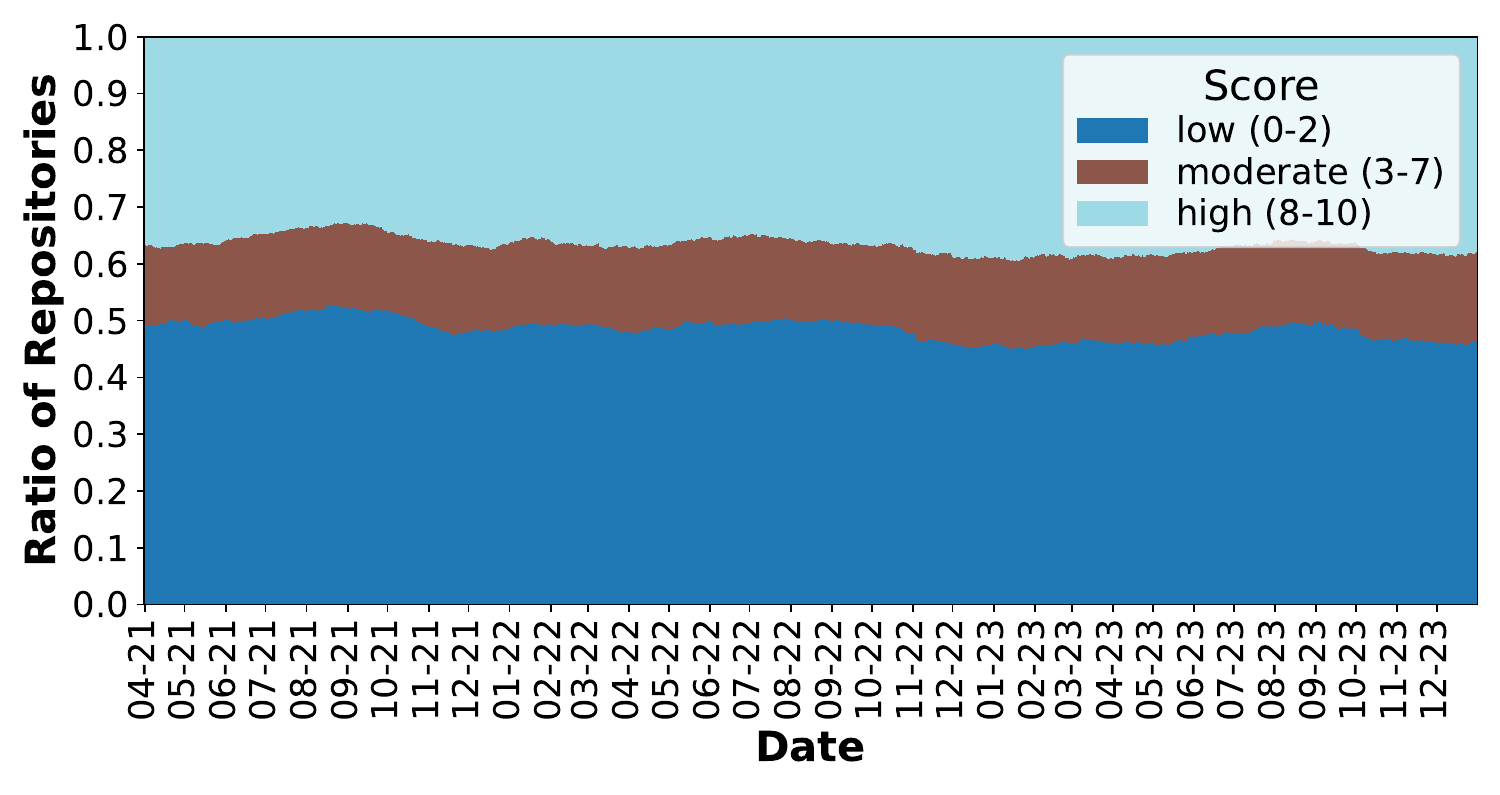}
    \caption{Temporal distribution of the bucketed \textit{Maintained} scores}
    \label{fig:rel_bucketed_Distribution_Score_Time}
\end{figure}

Figure~\ref{fig:rel_bucketed_Distribution_Score_Time} illustrates the temporal evolution of the three buckets between~2021 and~2023. Repositories with a \textit{low} \textit{Maintained} score constitute the largest share with~48\%, followed by~37\% classified as \textit{high} and~15\% as \textit{moderate}. Despite a remaining skew toward the extremes, the bucketed representation reduces the imbalance of the raw score by better representing mid-range activities, resulting in a clearer separation of maintenance levels. Therefore, this reformulation offers a stronger and more interpretable basis for subsequent forecasting tasks.

\subsubsection{Representation of the Target Variable as Trend Slope}
To capture not only the absolute values of the \textit{Maintained} score but also its dynamics over time, we derived the slope between consecutive monthly scores. This transformation quantifies the rate of change, reflecting to what extent a repository’s maintenance state improves, declines, or remains stable between two time steps. Consequently, the forecasting models can focus on temporal variations rather than static levels. 

\begin{figure}[h]
    \centering
    \includegraphics[width=0.85\textwidth, trim=2mm 3mm 2mm 2mm, clip]{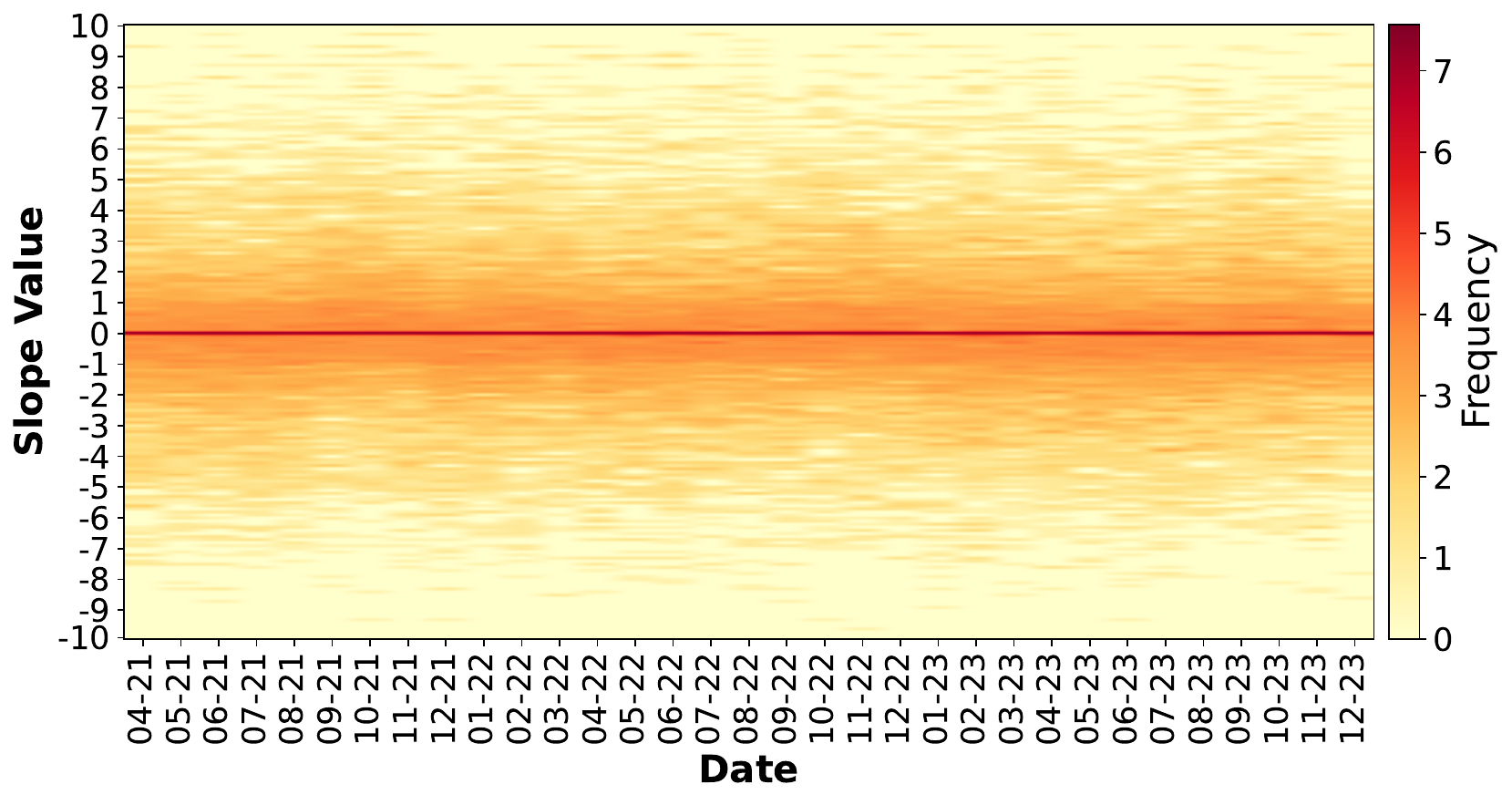}
    \caption{Temporal distribution of the trend slopes (log-scale)}
    \label{fig:Frequency_Heatmap_of_slope_values}
\end{figure}

Figure~\ref{fig:Frequency_Heatmap_of_slope_values} shows the frequency distribution of trend slopes between consecutive 30-day averages of the \textit{Maintained} score. The values are plotted on a logarithmic scale to account for the skewed data distribution. The heatmap reveals that most trend slopes cluster around zero, indicating that these repositories maintain stable activity across months. Positive and negative slopes occur less frequently, representing periods of increasing or decreasing maintenance activity. However, the slope values reach close to the extremes of -10 and 10, reflecting that the dataset captures a broad range of possible changes in the \textit{Maintained} score---from rapid declines to strong positive~shifts.

\subsubsection{Representation of the Target Variable as Trend Type}
To further generalize the temporal behavior of maintenance activity we transformed the continuous trend slopes into categorical trend types: \textit{downward}, \textit{stable}, and \textit{upward}. This discretization reduces noise in the target variable and enables the models to focus on categorical shifts in maintenance dynamics rather than exact numerical magnitudes. Such categorical trends also provide more interpretable insights for practitioners.

\begin{figure}[h]
        \centering
        \includegraphics[width=0.76\textwidth, trim=2mm 3mm 2mm 2mm, clip]{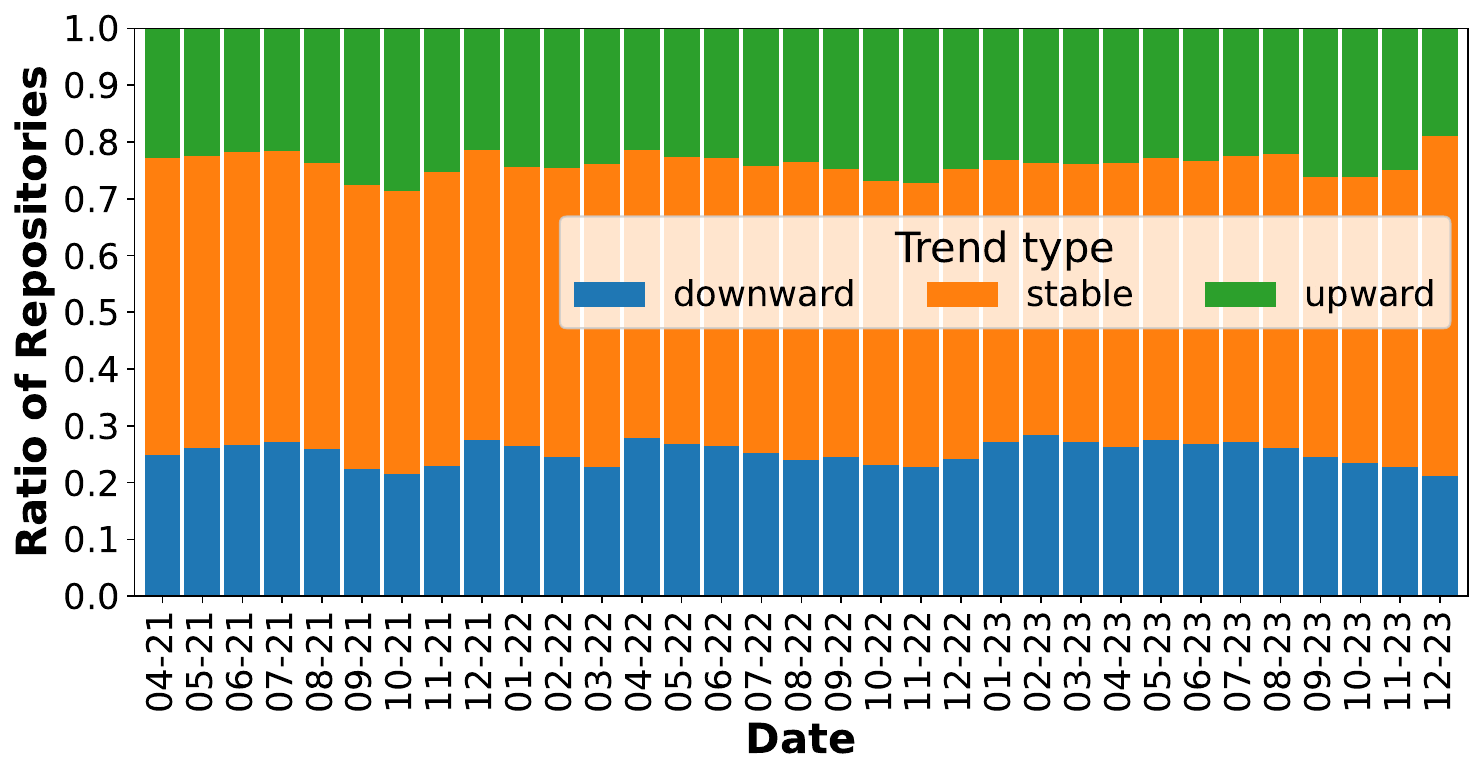}
        \caption{Temporal distribution of the trend types}
        \label{fig:Rolling_Trend_Classification_Ratios}
\end{figure}

Figure~\ref{fig:Rolling_Trend_Classification_Ratios} presents the temporal distribution of trend types using a 30-day sliding window. The \textit{stable} trend clearly dominates throughout the observation period, accounting for roughly half of the observations, while the \textit{downward} and \textit{upward} trends alternate at lower but comparable proportions. This pattern reflects the persistence of consistent maintenance activity across the majority of repositories, punctuated by occasional phases of decreasing or increasing engagement. Overall, this transformation simplifies the prediction task while preserving essential trend dynamics, thereby establishing a robust foundation for subsequent~forecasting.

\subsection{Model Development and Evaluation}
\label{sec:data_analsis}

\subsubsection{Model Training}
\label{sec:model_training}

We evaluate two model families to forecast the four representations of the target variable: a machine learning model (Random Forest, RF) and a deep learning model (Long Short-Term Memory, LSTM). To reduce seasonal effects, each model was retrained monthly (RF from scratch; LSTM incrementally) for a given training window length. The training data was constructed using a sliding-window, where each sample corresponds to a fixed-length sequence of past observations used to predict the \textit{Maintained} score---or a derived representation---for one or more subsequent months. All models were trained on identical inputs and targets to ensure comparability. Note, that the selected models do not represent state-of-the-art approaches in each family, but rather serve as well-established baselines to assess the feasibility of forecasting different representations of the \textit{Maintained} score.

\textbf{Machine Learning Approach.}
The RF regressor was implemented using the \texttt{sklearn} framework, which models non-linear relationships between temporal features and maintenance dynamics. Each multivariate sequence was flattened from its original $(n, t, f)$ structure into $(n, t \times f)$ to comply with the regressor’s tabular input format. The ensemble was trained with 100 estimators, unrestricted tree depth, and a fixed random seed to ensure reproducibility. By aggregating multiple decorrelated decision trees, the model captures non-linear interactions while mitigating overfitting. This provides a robust and computationally efficient benchmark against more complex sequential architectures such as LSTM models.

\textbf{Deep Learning Approach.}
We used the \texttt{TensorFlow/Keras} framework to implement the LSTM network, which directly models temporal dependencies within the multivariate sequences. The architecture consists of one or more recurrent layers with dropout-based regularization, followed by fully connected layers that map learned temporal representations to the target variable. Architectural depth, hidden dimensionality, and regularization strength were adapted to the specific prediction task (daily score prediction versus slope prediction). The network was trained using the Adam optimizer with a regression loss, mini-batch training, and early stopping to improve generalization. Prior to training, target values were clipped to $[0,10]$ for original mode and $[-10,10]$ for slope mode to enhance forecasting accuracy.

\subsubsection{Model Evaluation}
\label{sec:evaluation}

\textbf{Evaluation Setup:} Our evaluation systematically assesses the predictive performance of all model configurations while mitigating seasonal effects and ensuring comparability across forecasting setups. For every combination of target representation (raw \textit{Maintained} score, bucketed \textit{Maintained} score, trend slope, and trend type), model type (RF and LSTM), training window length, and forecasting horizon, models were evaluated using a sliding-window approach spanning twelve consecutive months. For slope-based forecasting, this corresponds to thirteen consecutive months of aggregated data in order to derive twelve month-to-month slope targets. In each iteration, a model was trained on a fixed-length historical training window and tested on the subsequent forecasting horizon. This process was repeated for all combinations of training window length (3 to 12 months) and forecasting horizon ranging from 1 to 6 months ahead, resulting in a comprehensive grid of temporal forecasting scenarios. To ensure a realistic forecasting setup and prevent data leakage, each sliding-window iteration used only information available up to the month being predicted. This design accounts for temporal non-stationarity and avoids bias from seasonal or calendar effects. Performance metrics were aggregated across all months and predictions, providing a temporally robust measure of model performance. To contextualize forecasting performance, we include a simple majority voting baseline that always predicts the most frequent class in the training window or the most frequent discretized value for regression targets. This baseline provides a lower-bound reference for assessing whether temporal models outperform dominant-class persistence. While alternative baselines (e.g., last-value or recency-weighted predictors) could also be considered, we focus on a frequency-based baseline to reflect the dominant class distribution observed in the data, as evidenced in Section~\ref{sec:time_series_anaylsis}.

\textbf{Performance Metrics:}
To ensure a consistent and directly comparable evaluation across all research questions, we focus primarily on \emph{accuracy} as the main performance metric in Section~\ref{sec:results}. This choice is motivated by the fact that all target representations are ultimately used in a discretized form to support decision-making (e.g., identifying maintenance levels or detecting changes in maintenance behavior), where correctly predicting the observed state or category is of primary practical relevance~\cite{hand2024selecting}. This requirement is directly captured by accuracy, which measures the proportion of correct predictions. Although RQ1 (raw \textit{Maintained} score) and RQ3 (trend slope) are formulated as regression tasks, their predictions are discretized prior to evaluation, enabling an unified accuracy-based assessment across continuous and categorical target representations. Performance values were aggregated across the monthly sliding-windows and across combinations of model type, target representation, training window length, and forecasting horizon. This aggregation enables a structured, multi-dimensional comparison of forecasting capability across methodological families and temporal contexts. Additional classification and regression metrics (e.g., MAE, $R^2$, F1-score) are provided in the replication package, enabling a more fine-grained analysis beyond the scope of this paper~\cite{replication_package}. Furthermore, in Section~\ref{sec:discussion}, we discuss confusion matrices for the two best-performing target representations to highlight misclassifications and near-miss errors.
\section{Results}
\label{sec:results}

\begin{figure}[ht]
    \centering
    \includegraphics[width=0.77\textwidth, trim=2mm 2mm 2mm 2mm, clip]{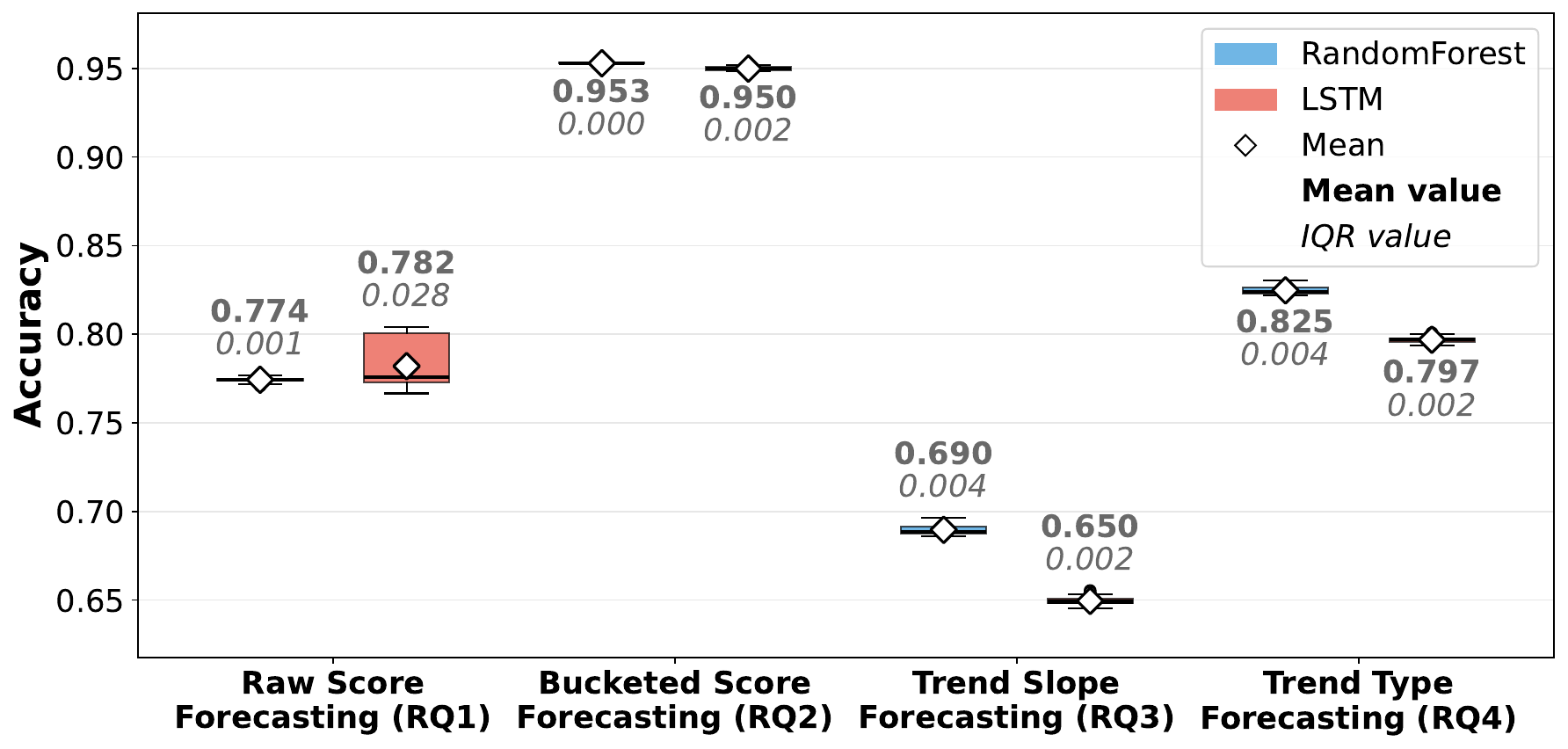}
    \caption{Aggregated accuracy across model families and target representations, averaged over training windows, forecasting horizons, and temporally shifted sliding-windows}

    \label{fig:model_performance}
\end{figure}

Figure~\ref{fig:model_performance} provides an aggregated overview of forecasting performance across all experimental configurations and serves as the basis for answering the four research questions (RQs). Each column corresponds to one forecasting task, reflecting a different representation of the target variable: raw \textit{Maintained} score forecasting, bucketed \textit{Maintained} score forecasting, trend slope forecasting, and trend type forecasting. Within each column, the two boxplots summarize the accuracy distributions achieved by the machine learning (Random Forest, RF) and deep learning (Long Short-Term Memory, LSTM) models. Each boxplot aggregates results across all training window lengths and forecasting horizons, as well as across twelve temporally shifted sliding-window evaluations to mitigate seasonal effects. This enables a comparison of model families and target representations while abstracting from individual parameter choices. Details about individual parameters are provided in the replication package~\cite{replication_package}.

\subsection{RQ1: Forecasting the Raw Score}
To address RQ1, we examine the first column in Figure~\ref{fig:model_performance}, which summarizes the accuracy achieved when forecasting the raw OpenSSF \textit{Maintained} score using Random Forest and LSTM models. Across all experimental configurations, the observed mean accuracies cluster in a narrow range of approximately 0.77--0.78, indicating that forecasting the raw \textit{Maintained} score is feasible for both model families. 
LSTM achieves the highest aggregated performance, with a mean accuracy of approximately 0.782, followed closely by the Random Forest with a mean accuracy of approximately 0.774. While the mean accuracies are comparable across models, differences emerge in terms of performance stability. Specifically, the interquartile ranges for Random Forest (IQR $\approx$ 0.001) are compact, indicating stable accuracy across different training window lengths and forecasting horizons. In contrast, the LSTM exhibits a wider interquartile range (IQR $\approx$ 0.028), reflecting higher variability across temporal configurations. The small performance differences between model families suggest that increased model complexity does not translate into improved predictive accuracy. These results indicate that the raw OpenSSF \textit{Maintained} score can be forecasted reliably---especially when using established machine learning techniques. Notably, the performance of all models clearly exceeds the majority voting baseline accuracy of 0.312, indicating that the models capture temporal maintenance patterns beyond dominant-score persistence.

\subsection{RQ2: Forecasting the Bucketed Score}
To answer RQ2, we examine the second column in Figure~\ref{fig:model_performance}, which summarizes the accuracy achieved when forecasting the bucketed OpenSSF \textit{Maintained} score using Random Forest and LSTM models. Compared to forecasting the raw score, both models achieve substantially higher mean accuracies of approximately 0.95, indicating that predicting broader maintenance activity levels is considerably more reliable. 
Among the evaluated approaches, the Random Forest achieves the highest aggregated performance, with a mean accuracy of approximately 0.953, while LSTM attains a slightly lower mean accuracy of approximately 0.950. The performance distributions of both models are tightly concentrated, with very small interquartile ranges (IQR $\approx$ 0.000 for Random Forest, IQR $\approx$ 0.002 for LSTM), reflecting consistently high accuracy across aggregated training window lengths and forecasting horizons. In contrast to RQ1, the reduced variability and uniformly high accuracy suggest that aggregating raw scores into broader maintenance categories simplifies the prediction task and mitigates sensitivity to individual experimental settings. Overall, these results demonstrate that the bucketed representation of the OpenSSF \textit{Maintained} score can be forecasted with high accuracy using machine learning and deep learning methods. Compared to a majority voting baseline accuracy of 0.465, all models provide a substantial improvement, confirming that bucketed maintenance levels are not predictable by class frequency alone.

\subsection{RQ3: Forecasting the Trend Slope}
To answer RQ3, we examine the third column in Figure~\ref{fig:model_performance}, which summarizes the accuracy achieved when forecasting the trend slope between consecutive OpenSSF \textit{Maintained} scores using Random Forest and LSTM models. Compared to forecasting the raw and bucketed \textit{Maintained} scores, predicting the trend slope is more challenging, as reflected by a lower overall accuracy of approximately 0.65--0.69 across all model families.
Among the evaluated approaches, the Random Forest achieves the highest aggregated performance, with a mean accuracy of approximately 0.690, while the LSTM attains a lower mean accuracy of approximately 0.650. The performance distributions show compact but model-dependent variability, with interquartile ranges of approximately 0.004 for Random Forest and 0.002 for LSTM. The consistently lower mean accuracy observed for the LSTM indicates reduced robustness across aggregated training window lengths and forecasting horizons. Overall, the lower mean accuracies compared to RQ2 suggest that forecasting the rate of change in maintenance activity is more difficult than predicting absolute maintenance levels. Although forecasting trend slopes remains challenging, all approaches outperform the majority voting baseline (accuracy 0.632), demonstrating added value from learned temporal dependencies.

\subsection{RQ4: Forecasting the Trend Type}
To answer RQ4, we examine the fourth column in Figure~\ref{fig:model_performance}, which summarizes the accuracy achieved when forecasting the trend type of the OpenSSF \textit{Maintained} score (downward, stable, upward) using Random Forest and LSTM models. Compared to forecasting the numerical trend slope in RQ3, predicting categorical trend types leads to substantially higher accuracy of approximately 0.80--0.83 across all model families. 
Among the evaluated approaches, the Random Forest achieves the highest aggregated performance, with a mean accuracy of approximately 0.825, while the LSTM attains a lower mean accuracy of approximately 0.797. The performance distributions of both models are compact, with small interquartile ranges (IQR $\approx$ 0.004 for Random Forest, IQR $\approx$ 0.002 for LSTM), indicating stable accuracy across aggregated training window lengths and forecasting horizons. Overall, the higher mean accuracies compared to RQ3 suggest that discretizing maintenance dynamics into categorical trend types simplifies the prediction task and mitigates some of the uncertainty associated with forecasting continuous trend slopes. Importantly, the achieved accuracies remain consistently above the majority voting baseline of 0.632, indicating that the models learn meaningful trend dynamics beyond majority-class stability.

\section{Discussion}
\label{sec:discussion}

\subsection{Comparative Difficulty of Forecasting Tasks}
Across the research questions, a clear hierarchy emerges regarding the forecastability of different representations of the OpenSSF \textit{Maintained} score. Tasks operating on higher abstraction levels, namely bucketed score~(RQ2) and trend types (RQ4), consistently achieve higher accuracy than forecasting raw scores (RQ1) or trend slopes~(RQ3). This gradient suggests that maintenance activity exhibits temporal regularities that are more easily captured when short-term noise and fine-grained numeric variability are abstracted away. This observation aligns with prior work emphasizing that forecasting performance depends strongly on how temporal patterns and data characteristics are represented and modeled~\cite{spiliotis2023time, li2024deep}.

\subsection{Model Complexity versus Predictive Benefit}
The results further indicate that increasing model complexity does not systematically improve forecasting performance. Across all target representations, Random Forest models achieve average performance comparable to---and in some cases slightly outperform---LSTM-based approaches. This observation is reinforced by the consistently small interquartile ranges observed for the machine learning models, which indicate stable performance across different training window lengths and forecasting horizons. In contrast, the LSTM models exhibit higher variability under certain configurations without achieving superior mean accuracy. These findings imply that simpler models offer a favorable trade-off between predictive performance, robustness, and computational cost. This is consistent with prior work showing that simpler machine learning approaches can perform competitively with more complex models for certain time series, depending on the data characteristics~\cite{romanov2023time}.

\subsection{Effect of Training Window Length and Forecasting Horizon}
The effect of the training window length and forecasting horizon is reflected in the variability of forecasting performance across experimental configurations, as quantified by interquartile ranges. Accuracy distributions are generally compact across target representations and models, indicating limited sensitivity to different combinations of training window lengths and forecasting horizons. Machine learning approaches exhibit consistently low interquartile ranges across tasks, while LSTM-based models show higher variability under certain configurations. Overall, the limited variability across most configurations suggests that reliable forecasting can already be achieved with small to medium training window lengths, without requiring extensive historical data. However, although forecasting performance remains robust even for longer forecasting horizons, such results should be interpreted with caution, as predictive uncertainty increases with horizon length and may affect practical applicability~\cite{mendis2024multivariate}.

\subsection{Class-level Error Patterns (RQ2 \& RQ4)}
\begin{figure}[ht]
    \centering
    \begin{subfigure}{0.49\textwidth}
        \centering
        \includegraphics[width=\textwidth, trim=0mm 0mm 0mm 0mm, clip]{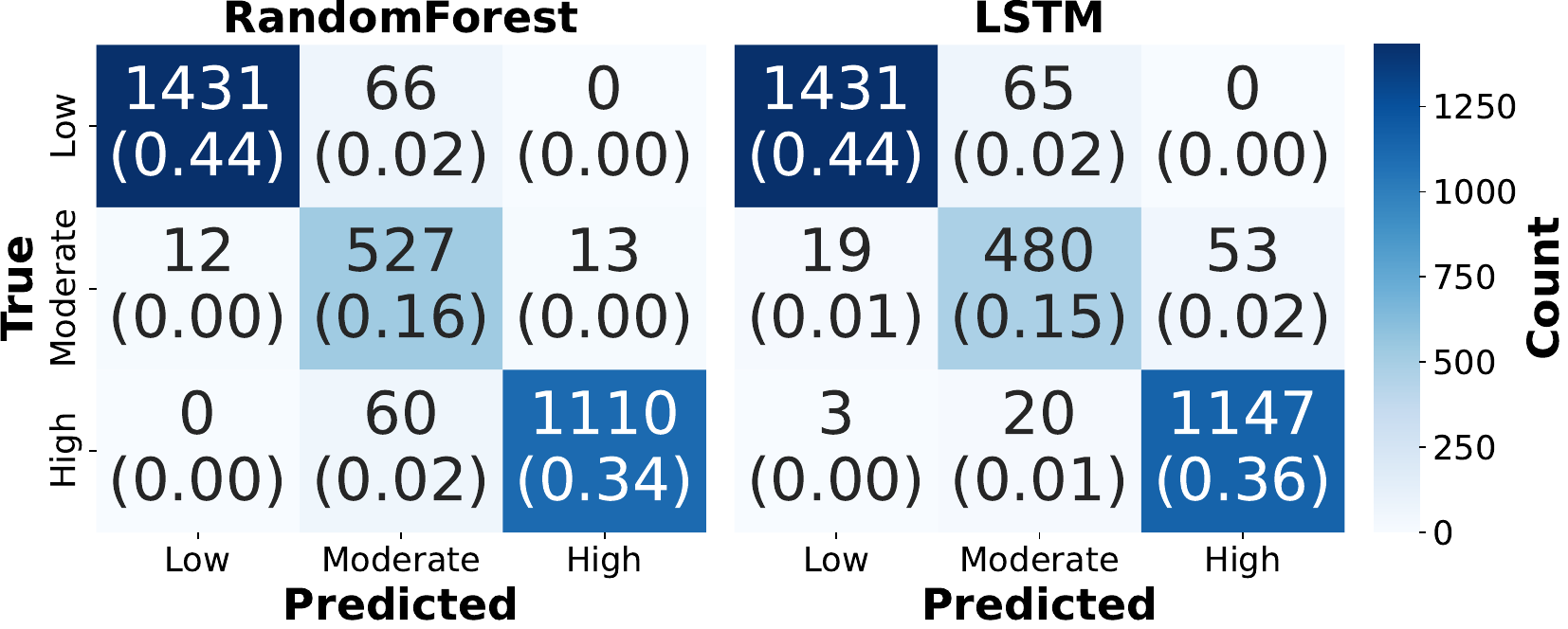}
        \caption{Bucketed Score (RQ2)}
        \label{fig:confusion_bucketed}
    \end{subfigure}
    \hfill
    \begin{subfigure}{0.49\textwidth}
        \centering
        \includegraphics[width=\textwidth, trim=0mm 0mm 0mm 0mm, clip]{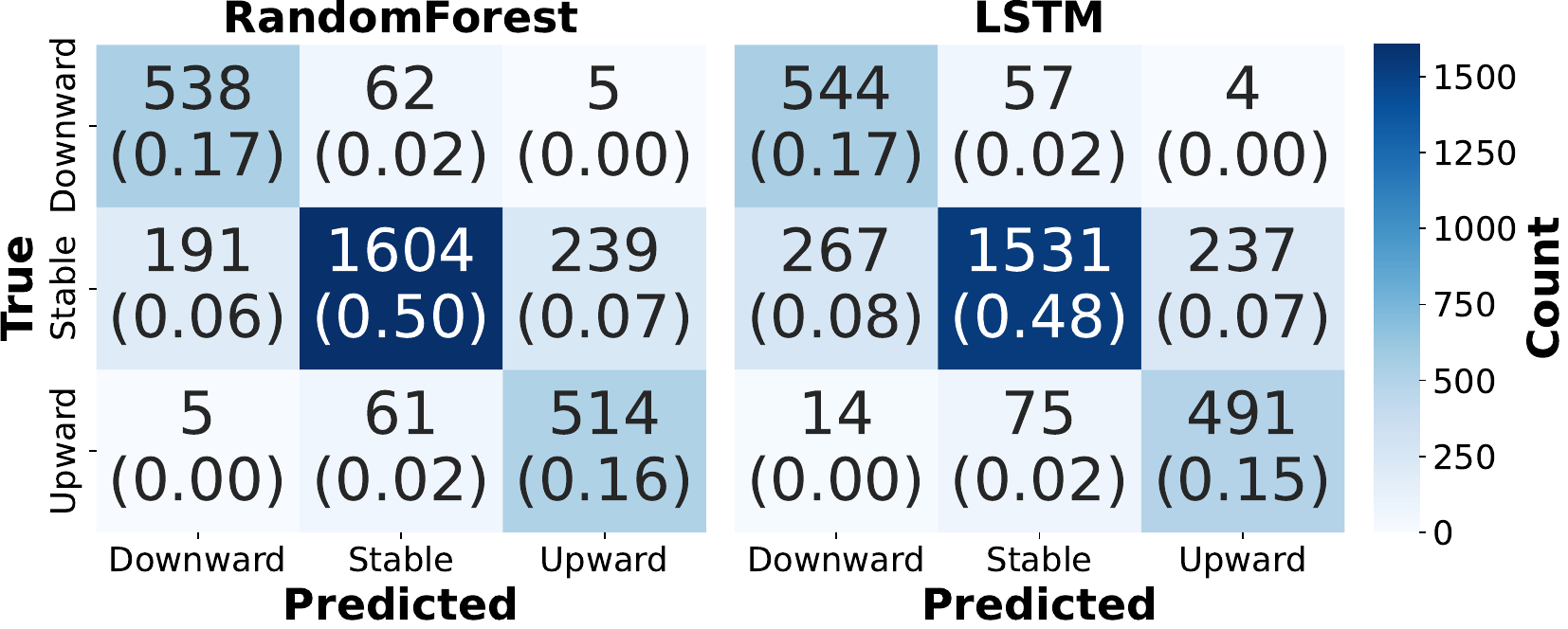}
        \caption{Trend Type (RQ4)}
        \label{fig:confusion_trend}
    \end{subfigure}
    \caption{Confusion matrices to assess misclassifications}
    \label{fig:confusion_matrices}
\end{figure}

Figure~\ref{fig:confusion_bucketed} and~\ref{fig:confusion_trend} show that the high accuracy for bucketed score~(RQ2) and trend type~(RQ4) forecasting is driven by class separability rather than majority-class bias. In the bucketed score setting, diagonal entries dominate, indicating reliable identification of low, moderate, and high maintenance levels, with misclassifications largely confined to adjacent buckets. Similarly, for trend type forecasting, the stable class is predicted with high reliability, and most errors occur between stable and downward or stable and upward trends. Direct confusion between extreme classes---low versus high or downward versus upward---is rare, suggesting that the models capture meaningful maintenance distinctions and directional signals.

\subsection{Practical Implications for Maintenance Assessment and Tooling}
Our findings suggest that forecasting abstracted representations of maintenance activity---namely bucketed scores and trend types---offers the most reliable insights. The observed error patterns indicate that even incorrect predictions tend to be close to the true state, which is critical for practitioner trust. Such forecasts can complement the OpenSSF Scorecard alongside its existing retrospective metrics by providing forward-looking signals. These signals support proactive dependency management and enable risk-based alerts for repositories with predicted declining maintenance. This addresses the lack of predictive capabilities in existing Scorecard-based analyses, which are typically limited to single point-in-time assessments~\cite{zahan2023openssf, mounesan2023exploring}.
Building on these capabilities, these forecasts can assist OSS maintainers in daily decision-making. Depending on the use case, some scenarios benefit from a joint assessment of current and predicted maintenance levels and trends, while others can be interpreted based on either dimension in isolation. For instance, predicted downward trends alone may indicate the need to allocate additional resources, prioritize issue resolution, or recruit contributors. In contrast, maintenance planning and release scheduling are particularly appropriate during periods of current and predicted high maintenance levels combined with stable or upward trend forecasts. In these cases, maintainers can schedule coordination-intensive tasks such as larger refactorings or dependency updates, or align releases with periods of sustained contributor activity. Moreover, integrating such forecasts into repository platforms could support code review and project oversight by highlighting OSS components at risk of declining maintenance activity, as well as helping developers select more sustainable dependencies.
The strong performance of comparatively simple models further lowers the barrier for adoption, making maintenance activity forecasting feasible for integration into real-world tooling. At the same time, such forecasts should be interpreted as probabilistic signals rather than guarantees, as over-reliance on predictions may create a false sense of safety. Thus, they are intended to complement continuous monitoring rather than replace~it.

\section{Threats to Validity}
Threats to validity are divided into the following four aspects~\cite{runeson2009guidelines}: 

\textbf{Construct validity:}
Some minor threats to construct validity arise from the selection of repositories. First, we only consider repositories explicitly linked on a PyPI library’s project page, potentially excluding forgotten links. However, prior work shows that most central PyPI libraries include a GitHub repository URL~\cite{tsakpinis2024analyzing}. Second, we argue that constructing a balanced training set by prioritizing the most central libraries improves practical applicability by focusing on the most impactful parts of the ecosystem, instead of using synthetically balanced repositories. Third, although the OpenSSF Scorecard supports GitLab, we focus exclusively on GitHub repositories, as only a small fraction of PyPI packages ($\approx2\%$) link to alternative hosts~\cite{tsakpinis2024analyzing}. Finally, library importance is computed without dependency version information, assuming always the latest version. While this may not fully reflect ecosystem dynamics, we expect its impact on PageRank scores to be negligible.

\textbf{Internal validity:} 
Although we do not examine causal relationships, internal validity may be affected by potential data leakage or inconsistent training procedures. These risks were mitigated through a consistent sliding-window design, identical data splits across models, and fixed random seeds to ensure reproducibility and prevent bias in performance estimation.

\textbf{External validity:} 
We focus on the PyPI ecosystem and GitHub as the code management system (CMS). Although the findings may not generalize to other ecosystems and CMSs, the methodology can be adapted to any publicly accessible ecosystem where library metadata include repository links, dependency information, and data required to compute a repository’s \textit{Maintained} score. In addition, because the study focuses on the top 1\% of PyPI libraries ranked by PageRank, the findings are most applicable to high-impact libraries, while generalizability to less central libraries remains an open question.

\textbf{Reliability:} 
Although PyPI and GitHub data are collected from publicly accessible sources, reproducibility is challenged by the dynamic nature of the PyPI ecosystem. Dependency information used to compute the library importance reflects a snapshot at the time of collection (November 2024), as PyPI does not provide historical dependency data. To address this, the collected data and analysis code are publicly available on Figshare~\cite{replication_package}. Additional risks from newly introduced, modified, or removed libraries, as well as updated or disabled repository links, can be mitigated by repeating the data collection. Nevertheless, if repository URLs remain unchanged and repositories persist, project metadata, issues, and commit histories can still be reproduced due to GitHub’s persistent hosting model.

\section{Conclusion and Future Work}

In this paper, we examined whether maintenance activities, as measured by the OpenSSF Scorecard \textit{Maintained} score, can be forecasted from historical repository data. Using GitHub repositories linked to one of the top 1\% most central PyPI libraries as measured by PageRank, we reconstructed \textit{Maintained} scores between 2021 and 2023 and evaluated machine learning and deep learning approaches using training windows of 3--12 months and forecasting horizons of 1--6 months.
Our results show that the \textit{Maintained} score exhibits strong temporal predictability. Forecasting the raw \textit{Maintained} score (ranging from 0 to 10) achieves stable performance across all model families, with mean accuracies around 0.77--0.78, indicating that future maintenance levels can be estimated reliably from past activity. Aggregating raw scores into broader maintenance categories substantially improves performance: bucketed score forecasting---grouping scores into \textit{low} (0--2), \textit{moderate} (3--7), and \textit{high} (8--10)---achieves mean accuracies around 0.95 across all models, while trend type forecasting (downward, stable, upward) reaches accuracies of approximately 0.80--0.83. In contrast, forecasting numerical trend slopes proves more challenging, with mean accuracies around 0.65--0.69, reflecting the difficulty of predicting short-term fluctuations in maintenance behavior.
Across all target representations, Random Forest models perform on average comparably to, and sometimes slightly better than, LSTM models, while exhibiting lower variability and computational cost, indicating that the temporal dynamics of the \textit{Maintained} score can be effectively modeled using simpler and more interpretable techniques. Overall, this study demonstrates that the OpenSSF \textit{Maintained} metric can be extended beyond retrospective assessment to support forward-looking analysis. Depending on the use case, practitioners can trade granularity for robustness: raw score forecasts enable fine-grained analysis, while bucketed scores and trend types provide highly accurate and interpretable signals for proactive risk assessment.

Future work can build on these findings in three directions. First, a deeper investigation of prediction errors is needed. Analyzing misclassifications in the confusion matrices---especially near-miss errors between adjacent classes---may reveal underlying causes and inform targeted feature engineering and model refinement. Second, future studies should explore personalized forecasting models trained solely on the historical data of individual repositories. Such models may better capture repository-specific maintenance rhythms and governance practices, complementing the global models used in this study. Third, integrating maintenance activity forecasting into practitioner-facing tooling, such as the OpenSSF Scorecard or dependency analysis platforms, would enable continuous, forward-looking assessment of repository health and support more informed dependency management decisions.
\section{Data Availability}
\label{sec:data_availabaility}
The dataset supporting this study, along with the analysis artifacts, is openly available on Figshare under a CC-BY 4.0 license~\cite{replication_package}. The replication package contains all materials necessary to reproduce the reported results.

\bibliography{research_paper}

\end{document}